\documentclass[11pt]{article}
\usepackage{graphicx}
\usepackage[utf8]{inputenc}
\usepackage[T1]{fontenc}
\usepackage{cite}

\def\nn{\nonumber}

\def\arcsinh{\mathop{\mbox{arcsinh}}}
\def\diag{\mathop{\mbox{diag}}}
\def\eff{\mbox{\footnotesize\it eff}}

\title{On dark stars, Planck cores\\ and the nature of dark matter}
\author{Igor Nikitin\\
Fraunhofer Institute for Algorithms and Scientific Computing\\
Schloss Birlinghoven, 53757 Sankt Augustin, Germany\\
\\
igor.nikitin@scai.fraunhofer.de
}
\date{}

\begin{document}
\maketitle

\begin{abstract}
Dark stars are compact massive objects, described by Einstein gravitational field equations with matter. The type we consider possesses no event horizon, instead, there is a deep gravitational well with a very strong redshift factor. Observationally, dark stars can be identified with black holes. Inside dark stars, Planck density of matter is reached, Planck cores are formed, where the equations are modified by quantum gravity. In the paper, several models of dark stars with Planck cores are considered, resulting in the following hypothesis on the composition of dark matter. The galaxies are flooded with low-energetic radiation from the dark stars. The particle type can be photons and gravitons from the Standard Model, can also be a new type of massless particles. The model estimations show that the extremely large redshift factor $z\sim10^{49}$ and the emission wavelength $\lambda_0\sim10^{14}$m can be reached. The particles are not registered directly in the existing dark matter experiments. They come in a density sufficient to explain the observable rotation curves. The emission has a geometric dependence of density on radius $\rho\sim r^{-2}$, producing flat rotation curves. The distribution of sources also describes the deviations from the flat shape. The model provides a good fit of experimental rotation curves. Outbreaks caused by a fall of an external object on a dark star lead to emission wavelength shifted towards smaller values. The model estimations give the outbreak wavelength $\lambda\sim1$m compatible with fast radio bursts. The paper raises several principal questions. White holes with Planck core appear to be stable. Galactic rotation curves in the considered setup do not depend on the matter type. Inside the galaxy, dark matter can be of hot radial type. At cosmological distances, it can behave like the cold uniform type.
\end{abstract}


\begin{figure}
\centering
\includegraphics[width=\textwidth]{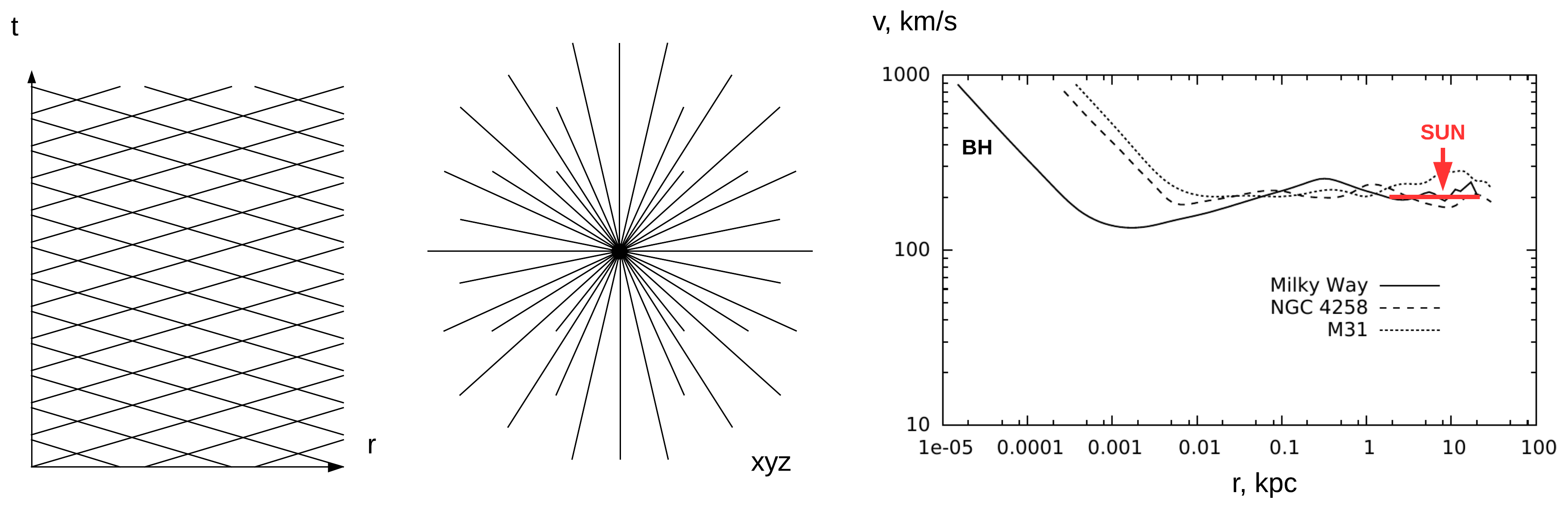}
\caption{On the left and in the center: an RDM-star -- a black hole, coupled to radial flows of dark matter. On the right: experimental rotation curves for three galaxies. Image from~\cite{1701.01569}, data from \cite{VR2}.}
\label{p1f1}
\end{figure}

\section{Introduction}

Dark stars, also known as quasi black holes, boson stars, gravastars, fuzzballs, are solutions of general theory of relativity, which first follow the Schwarz\-schild profile and then are modified. Outside they are similar to black holes, inside they are constructed differently, depending on the model of matter used. An overview of these models can be found in the paper by Visser et al. ``Small, dark and heavy: but is this a black hole?'' \cite{0902.0346}. A recent advance has been reported by Holdom and Ren in their paper ``Not quite a black hole'' \cite{1612.04889}. Our contribution to this family are RDM-stars \cite{1701.01569}, quasi black holes coupled to Radial Dark Matter. A typical configuration of an RDM-star is shown on Fig.\ref{p1f1} on the left and in the center. It is a stationary solution, including  T-symmetric superposition of ingoing and outgoing radially directed flows of dark matter. 

An RDM-star can be used as the simplest model of a spiral galaxy. In the limit of weak gravitational fields, the dark matter flows radially converging towards the center of the galaxy produce a typical geometric dependence of mass density on the radius $\rho\sim r^{-2}$, which corresponds to constant orbital velocity $v=Const$, flat rotation curve. It is a qualitatively correct behavior for many experimental rotation curves at large distances, see Fig.\ref{p1f1} on the right. We will show that a distribution of RDM-stars in the galaxy also allows to describe correctly the deviations of rotation curves from the flat shape. The model of RDM-stars fits very well the experimental rotation curves by Sofue et al. \cite{VR2,0811.0859,0811.0860,1110.4431,1307.8241} and Salucci et al. \cite{9502091,9503051,9506004,0703115,1609.06903}. 

In strong gravitational fields, RDM-stars behave interestingly. First of all, the event horizon, typical for real black holes, is erased. Instead, a deep gravitational well is formed, where the values of the redshift become enormously large. As a result, for an external observer the star looks black, like a real black hole. Simultaneously, the mass density increases rapidly, reaching and exceeding the Planck value.

This is where Planck stars come into play. This model is based on the calculations in quantum loop gravity, performed for a scalar field cosmology by Ashtekar et al. \cite{0602086,0604013,0607039}. According to these calculations, the mass density has a quantum correction: $\rho_X = \rho (1-\rho/\rho_c)$, where the critical density $\rho_c\sim\rho_P$ is of the order of Planck value, $\rho$ is the nominal density before the correction and $\rho_X$ is the effective density participating in Einstein field equations. As a result of this correction, $\rho = \rho_c$ corresponds to $\rho_X= 0$, at the critical density the gravity is effectively switched off, while $\rho > \rho_c$ corresponds to $\rho_X< 0$, in excess of critical density the effective negative mass appears (exotic matter), with gravitational repulsion (quantum bounce phenomenon). In the Planck star model by Rovelli, Vidotto \cite{1401.6562}, Barceló et al. \cite{1409.1501}, a collapse of a star leads to the quantum bounce, is replaced by extension, as a result, the black hole turns white. 

In this paper we consider a stationary version of a Planck star, stabilized under the pressure of the external matter (Planck core). We will consider two stationary spherically symmetric models with a Planck core in the center. The subject is related to the stability of white holes, earlier investigated in papers by Ori and Poisson \cite{OriPoisson}, Eardley \cite{Eardley}, Zel'dovich, Novikov and Starobinskij \cite{ZNS}. It is also related to the origin of fast radio bursts and gives an unusual viewpoint on the nature of dark matter.

The paper is organized as follows. In Section~2 the model of RDM-stars is considered. The main computations have been performed in the author's paper \cite{dm_stars}, here a short overview of the results is given. In Section~3 the model of TOV-stars with Planck core is presented. In Section~4 the nature of dark matter according to the considered models is discussed. A theoretically interesting question on stability of white holes is considered in the Appendix.

\section{RDM-stars with Planck core}

\paragraph*{RDM-stars and rotation curves of galaxies.} RDM-star geometry can be used as a simplest model of dark matter distribution in spiral galaxies. Let us consider dark matter flows radially converging towards the center of a galaxy, displayed on Fig.\ref{p1f1} center, in the limit of weak gravitational fields. The one-line calculation
\begin{eqnarray}
&&\rho_{dm}\sim r^{-2},\ M_{dm}\sim r,\ v^2=GM_{dm}/r=Const \label{oneline}
\end{eqnarray}
evaluates mass density, enclosed mass function and orbital velocity of stars. Adding a concentrated mass in the center, $v^2=GM_0/r+Const$, the rotation curve described by the sum of Keplerian and constant terms can be obtained. The real rotation curves, displayed on Fig.\ref{p1f1} right, possess a similar structure, with Keplerian behavior at small distances and flat shape at large distances. The red line shows a segment 2-20kpc where the rotation curve for the Milky Way can be considered as approximately flat, with the Sun position at 8kpc. These plots show that the real rotation curves deviate from a simple sum of Keplerian and constant terms, revealing additional structures, oscillations. On the other hand, the model with a single RDM-star in the center of the galaxy is also a simplification. Further we consider a model of distributed RDM-stars, able to capture the additional structures. Then we perform a calculation in the limit of strong gravitational fields to analyze the interior structure of an RDM-star.

The detailed description of rotation curves in the RDM-model is based on two assumptions: (1) all black holes are RDM-stars; (2) their density is proportional to the concentration of the luminous matter in the galaxy. As a result, the dark matter mass density can be represented by the integral
\begin{eqnarray}
&&\rho_{dm}(x)=\int d^3x'\, b(|x-x'|)\, \rho_{lm}(x'),\ b(r)=1/(4\pi L_{KT})/r^2. \label{KT1}
\end{eqnarray}
Here $\rho_{lm}$ is the density of luminous matter, the kernel $b(r)$ represents a contribution of a single RDM-star and $L_{KT}$ is a parameter of length dimension, regulating a coupling between the dark and the luminous matter. This form of coupling has been proposed earlier in a context of a different model in works by Kirillov and Turaev \cite{0202302,0604496}. The physical meaning of the $L_{KT}$ parameter is the radius at which the enclosed mass of dark matter equals to the mass of the luminous matter, to which it is coupled: $ M_ {dm} (L_ {KT}) = M_ {lm} $. 

\begin{figure}
\begin{center}
\includegraphics[width=\textwidth]{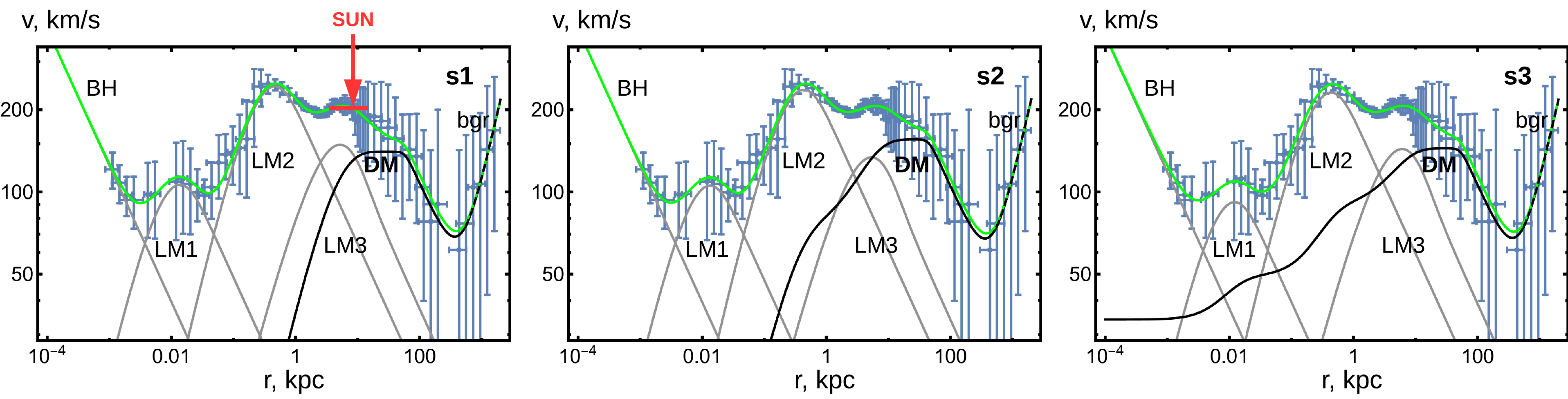}
\end{center}
\caption{Detailed rotation curve for Milky Way, fitted by RDM-model. Blue points with error bars -- data from \cite{1307.8241}. Green curve -- fit by RDM-model from \cite{dm_stars} for three coupling scenarios (s1-3). Contributions of different galactic structures are also shown.}\label{p1f2}
\end{figure}

\paragraph*{The detailed rotation curve of Milky Way,} known also as Grand Rotation Curve (GRC), has been constructed on the basis of various experimental data by Sofue et al. \cite{VR2,0811.0859,0811.0860,1110.4431,1307.8241}. This curve is presented on Fig.\ref{p1f2} by data points with errors. Here one can see several structures, including Keplerian contribution of the central black hole (BH), inner and outer bulges (LM1,2), galactic disk (LM3), followed by dark matter contribution (DM) and background outer part (bgr). The red line with the marked Sun position represents the same 2-20kpc approximately flat interval as on the previous figure. This part appears to be relatively small due to a much larger range of distances involved in the analysis.

In paper \cite{1307.8241}, the distribution of luminous matter in the bulges is described by {\it exponential spheroid model}, representing the mass density by an exponent $\rho_{lm}\sim\exp(-r/a)$. For the galactic disk {\it Freeman's model} \cite{freeman} is used, with the surface mass density described by similar exponent $\rho_{lm}\sim\delta(z)\exp(-r/R_D)$. Taking these distributions, the integral (\ref{KT1}) and the resulting rotation curve $v(r)$ can be evaluated analytically. The lengthy explicit expressions per every structure are given in \cite{dm_stars}, also used there as basis functions for the fitting procedure. For stability of the fit, the relative coupling of dark matter to different structures of luminous matter has been fixed as shown in Table~\ref{p1tab1}. The $\lambda$-constants are used as multiplicative factors to integrals (\ref{KT1}). Since the different galactic structures may possess a different density and different population of black holes, we can select different coupling constants for them. This procedure is equivalent to a readjustment of the corresponding $L_{KT}$-parameters, while we prefer to use a single $L_{KT}$-parameter and adjust the individual couplings by relative $\lambda$-factors. Three scenarios have been considered in Table~\ref{p1tab1}, the first one assigns all dark matter coupling to the galactic disk, the second one introduces equal coupling among all structures, the third one describes a prevailing coupling for the central structures.

The result of the fit is shown by curves on Fig.\ref{p1f2}. The green line represents the total rotation curve, a quadratic sum over all structures. It has almost the same shape for all three scenarios. Also, the separated contributions of different structures are shown. They depend on the scenario, e.g., the third scenario with prevailing dark matter coupling to central structures also shows a considerable contribution of dark matter in the center. Table~\ref{p1tab2} presents the obtained fitting parameters -- the total masses and geometric sizes of the structures. In the considered modeling, the dark matter halo is sharply cut at the radius $R_{cut}$, further providing Keplerian fall of the outer part of the rotation curve, followed by its linear increase due to the uniform background density. Interestingly, the parameters, characterizing the outer part of the rotation curve, the total mass of dark matter halo $M_{dm}(R_{cut})$ and the background density $\rho_{bgr}$, appear to be approximately the same for the considered three scenarios.

\begin{table}
\begin{center}{\footnotesize
\caption{GRC fit: coupling coefficients for 3 scenarios}\label{p1tab1}

~

\def\arraystretch{1.1}
\begin{tabular}{|c|c|c|c|}
\hline
$\lambda_{KT}$&s1&s2&s3\\
\hline
 $\lambda_{smbh}$& 0& 1& $10^3$ \\
 $\lambda_1$& 0& 1& $10^2$ \\
 $\lambda_2$& 0& 1& 2 \\
 $\lambda_{disk}$& 1& 1& 1 \\
\hline

\end{tabular}

}\end{center}
\end{table}

\begin{table}
\begin{center}{\footnotesize
\caption{GRC fit: the results*}\label{p1tab2}

~

\def\arraystretch{1.1}
\begin{tabular}{|c|c|c|c|}
\hline
$par$&s1&s2&s3\\
\hline
 $M_{smbh}$& $3.6\times 10^6$ & $3.6\times 10^6$ & $3.2\times 10^6$ \\
 $M_1$& $5.5\times 10^7$ & $5.2\times 10^7$ & $3.6\times 10^7$ \\
 $a_1$& $0.0041$ & $0.0039$ & $0.0036$ \\
 $M_2$& $9.7\times 10^9$ & $8.6\times 10^9$ & $8.2\times 10^9$ \\
 $a_2$& $0.13$ & $0.13$ & $0.13$ \\
 $M_{disk}$& $3.2\times 10^{10}$ & $2.7\times 10^{10}$ & $3.5\times 10^{10}$ \\
 $R_D$& $2.4$ & $2.5$ & $2.8$ \\
 $L_{KT}$& $7.0$ & $6.3$ & $12.0$ \\
 $R_{cut}$& $58$ & $45$ & $53$ \\
  $M_{dm}(R_{cut})$& $2.7\times 10^{11}$ & $2.5\times 10^{11}$ & $2.6\times 10^{11}$ \\
$\rho_{bgr}$& $646$ & $653$ & $649$ \\
\hline
\end{tabular}

* masses in $M_\odot$, lengths in $kpc$, density in $M_\odot/kpc^3$

}\end{center}
\end{table}

\begin{figure}
\begin{center}
\includegraphics[width=\textwidth]{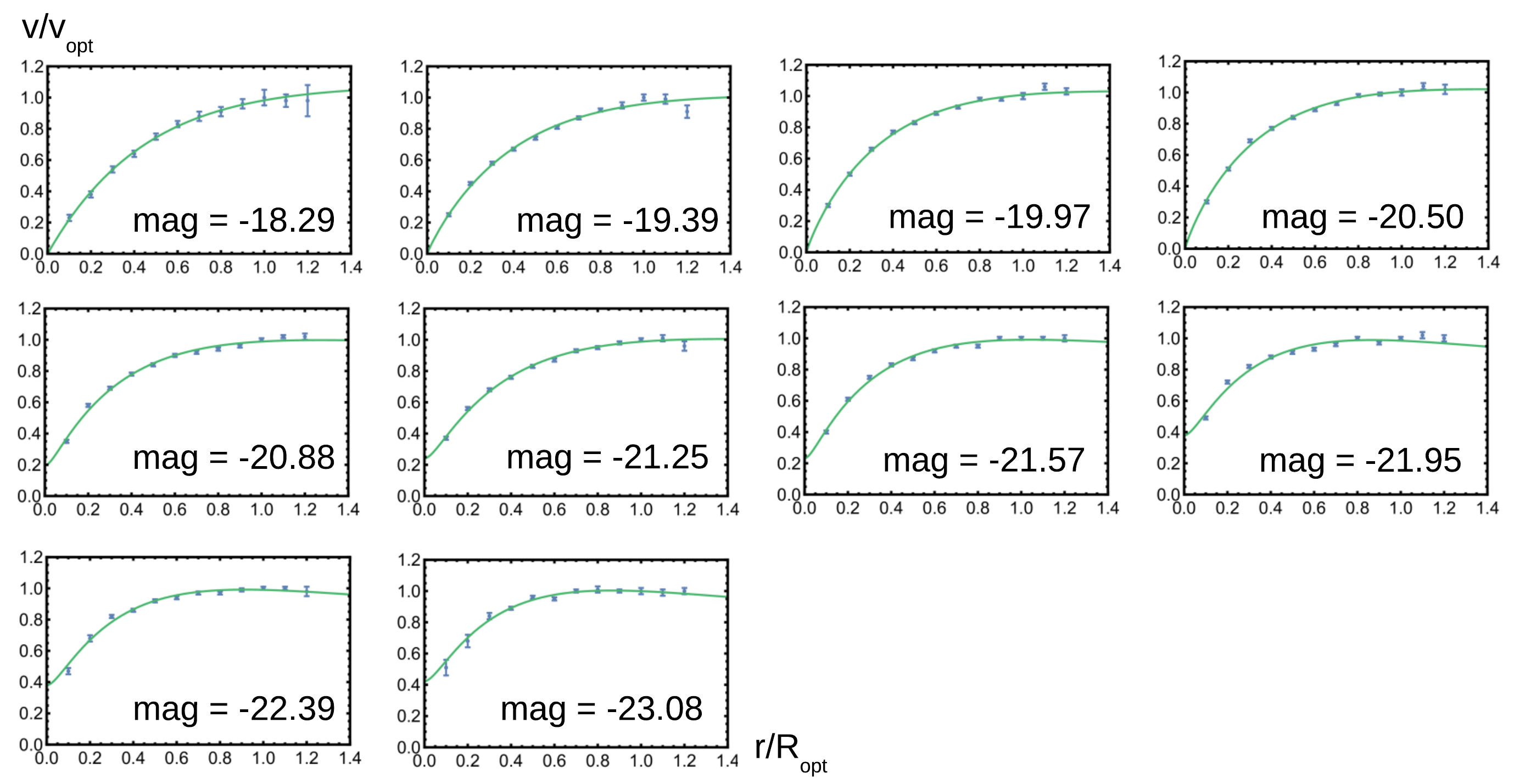}
\end{center}
\caption{Other galaxies: universal rotation curve, fitted by RDM model. The points with error bars -- data from \cite{9502091}. Green curves -- fit by RDM-model from \cite{dm_stars}. The data and the fits for different luminosity bins $mag$ are separated.}\label{p1f3}
\end{figure}

\paragraph*{Other galaxies} can be modeled with a concept of a Universal Rotation Curve (URC) introduced by Salucci et al. \cite{9502091,9503051,9506004,0703115,1609.06903}. It represents averaged experimental rotation curves of more than 1000 galaxies. Before averaging, the galaxies are subdivided to bins over the magnitude $mag$ and the curves $v(r,mag)$ are normalized to the values at optical radius: $v/v_{opt}$, $r/R_{opt}$. Here, $v_{opt}=v(R_{opt})$ and the optical radius of the galaxy $R_{opt}=3.2R_D$ is defined as a distance, under which 83\% of the luminous mass is located. The averaging smooths the individual features of the curves, their local minima and maxima. The resulting experimental curves appear to be more smooth and are shown by points with errors on Fig.\ref{p1f3}.

On these plots, the radius and velocity are presented in a linear scale, rather than the logarithmic one used in previous plots. As a result, the earlier described central structures are shrinked to a single unresolved central contribution. The modeling is accordingly simplified, preserving only the central and the disk contributions. The basis functions are explicitly written in \cite{dm_stars}, the result of the fit is presented by green curves on Fig.\ref{p1f3}. 

The presented plots show that the model of distributed RDM-stars, based on the Newtonian weak field limit and the proportionality assumptions above, provides a good fit of the experimental rotation curves, for both GRC and URC types.

\begin{figure}
\begin{center}
\includegraphics[width=\textwidth]{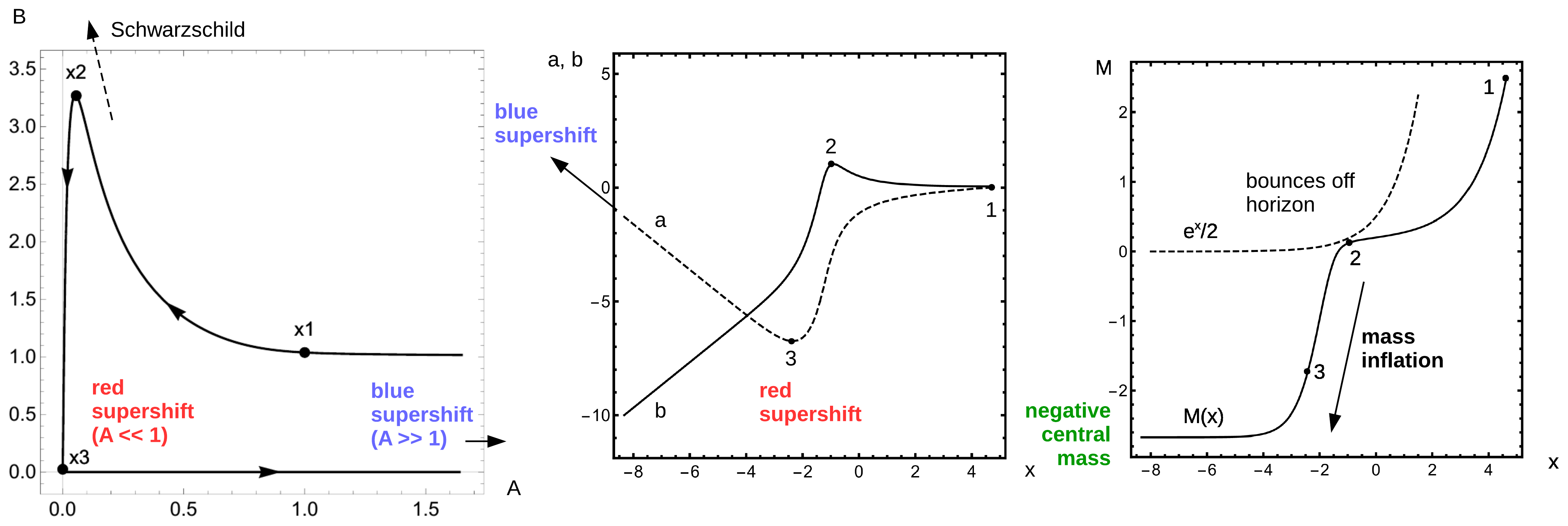}
\end{center}
\caption{RDM-star model in strong fields. A typical solution in different coordinates (see text).}\label{p1f4}
\end{figure}

\paragraph*{RDM-star model in strong fields.} The system to solve is combined from Einstein gravitational field equations and geodesic equations:
\begin{eqnarray}
&G^{\mu\nu}=8\pi G/c^4\cdot T^{\mu\nu},\ 
u^\nu\nabla_\nu u^\mu=0,\ \nabla_\mu\rho u^\mu=0.\label{eq_efe}
\end{eqnarray}
We consider the model with T-symmetric non-interacting superposition of ingoing and outgoing flows of dark matter. Therefore, geodesic equations can be applied separately for every flow, described by velocity field $u^\mu$ and intrinsic mass density $\rho$. A static spherically symmetric metric is chosen:
\begin{eqnarray}
&ds^2=-A dt^2+B dr^2+D r^2(d\theta^2+\sin^2\theta\; d\phi^2),\label{ds2}
\end{eqnarray}
where the profile $A(r)>0$ describes the redshift and time delay effects, $B(r)>0$ measures geometric deformation in radial direction. $D=1$ can be put by convention, so that $r$ is aerial radial coordinate, the area of $r$-sphere is $4\pi r^2$. Time $t$ is measured by the clock of a distant observer, where $A\to1$ can be set. Energy-momentum tensor is taken in a form
\begin{eqnarray}
&T^{\mu\nu}=\rho(u_+^\mu u_+^\nu+u_-^\mu u_-^\nu),\ u_\pm=(\pm u^t,u^r,0,0),
\end{eqnarray}
a sum of T-symmetric radial flows of non-interacting dust matter.

The equations (\ref{eq_efe}) have been solved in \cite{1701.01569,dm_stars}. Geodesic equations possess analytical solution
\begin{eqnarray}
&4\pi\rho=c_1/\left(r^2u^r\sqrt{AB}\right),\label{eq_geode}\\
&u^t=c_2/A, \ u^r=\sqrt{c_2^2+c_3A}/\sqrt{AB},\nn
\end{eqnarray}
in $G=c=1$ normalization. The integration constants $c_{1-3}$ will be considered in details later. The Einstein equations have a form
\begin{eqnarray}
&rA'=-A+AB+ 4c_1B\sqrt{c_2^2 + c_3A},\label{eq_rdm}\\
&rB'=B/A\left(A-AB+4c_1c_2^2B/\sqrt{c_2^2 + c_3A}\right),\nn
\end{eqnarray}
they can be solved numerically. The typical solution is shown on Fig.\ref{p1f4} left, in $(A,B)$-coordinates. Initially, near the point $x_1$, the curve has a hyperbolic form, typical for Schwarzschild solution. The difference starts near the point $x_2$, where the Schwarzschild solution goes to infinity, the event horizon is formed. In the considered solution, the dark matter acts like a barrier, preventing the formation of the horizon. The solution then goes rapidly towards very small values of $A$ and $B$, where it exhibits a strong redshift and possesses a small proper length. Then the solution goes to large values of $A$, a strong blueshift. The same solution is shown in the central part of this figure, in logarithmic coordinates, and on the right part, presenting a Misner-Sharp enclosed mass function:
\begin{eqnarray}
&&x=\log r,\ a=\log A,\ b=\log B,\ M=(1-B^{-1})\,r/2.\label{transf}
\end{eqnarray}
In these coordinates the equations obtain the form more convenient for a numerical solution
\begin{eqnarray}
&&a'_x=-1+e^b+c_4e^{b-a}\sqrt{1 + c_5e^a},\label{eq_dadx}\\
&&b'_x=1-e^b+c_4e^{b-a}/\sqrt{1 + c_5e^a},\label{eq_dbdx}\\
&&c_4=4c_1c_2,\ c_5=c_3/c_2^2.
\end{eqnarray}
The convenience follows from the resolution of singularities, typical for polynomial formulation, so that the resulting equations can be easily solved, e.g., by {\it Mathematica}~~{\tt NDSolve} algorithm. Also each term in these equations has a clearly defined range of domination, so that normally only one term in the equation is active. This simplifies the asymptotic analysis of the system.

The behavior of the mass function on Fig.\ref{p1f4} right shows that the solution is bounced off horizon line $M=r/2$, then falls very rapidly. This fall is related to the phenomenon of mass inflation, described in the paper by Hamilton and Pollack \cite{0411062}. There is a positive feedback loop in black hole solutions with counterstreaming matter flows: (1) increasing energy of the crossing flows leads to (2) increasing pressure, that leads to (3) increasing gravity, that leads again to (1). As a result, an accumulation of very large mass in the counterstreaming region happens. For the considered solutions, the function $M(r)$ decreases with decreasing $r$. To explain this property, one can imagine spherical shells of positive mass consequentially removed from the star. Finally, the mass arrives to a negative central value, a concentrated negative mass. It corresponds to the well known Schwarzschild singularity of naked type and explains the appearance of a blueshift region in the solution. On the other hand, the singularity is coated in a massive shell appearing due to the mass inflation phenomenon, so that the total mass of the system remains positive. Also, below we will introduce a quantum gravity cutoff in the model, which will remove the naked singularity with most of surrounding structures.

The integration constants $c_{1,2}>0$, while $c_3=u_\mu u^\mu=-1,0,+1$ can take three discrete values, corresponding to the type of dark matter particles: massive, null or (theoretically) tachyonic. Interestingly, the solution in strong fields ($A\ll1$) depends on the matter type very weakly, since the corresponding term $c_3A$ in the equations becomes small. Solution in weak fields ($A\sim1$) depends, at first, on the parameter $c_5$ that defines asymptotic radial velocity of the dark matter: for $c_5<-1$, the massive radial flow has a turning point, the matter cannot escape; $c_5>-1$, possible for all matter types, the matter can escape to large distances, the case further considered:
\begin{eqnarray}
&&c_6=c_4\sqrt{1 + c_5},\ c_7=c_4/\sqrt{1 + c_5},\ \epsilon=(c_6+c_7)/2. 
\end{eqnarray}
The parameter $\epsilon$ defines an asymptotic gravitating density $\rho_{grav}=\rho_{\eff}+p_{\eff}$. The effective density and pressure, produced by counterstreaming dark matter flows, are defined as components of energy-momentum tensor $T_\mu^\nu=\diag(-\rho_{\eff},p_{\eff},0,0)$, where
\begin{eqnarray}
&&\rho_{\eff}=c_4/(8\pi r^2A)/\sqrt{1+c_5A},\ p_{\eff}=c_4/(8\pi r^2A)\cdot\sqrt{1+c_5A}.\label{rhopeff0}
\end{eqnarray}
In the weak field limit $A\sim1$ we obtain $\rho_{grav}=\epsilon/(4\pi r^2)$, $M_{grav}=\epsilon r$, as in (\ref{oneline}). This makes $\epsilon$ a directly measurable parameter, in physical units $\epsilon=(v/c)^2$, where $v$ is the orbital velocity of stars at large distances from the galaxy center, for Milky Way $v\sim200$km/s, $\epsilon\sim4\cdot10^{-7}$.

\begin{figure}
\begin{center}
\includegraphics[width=\textwidth]{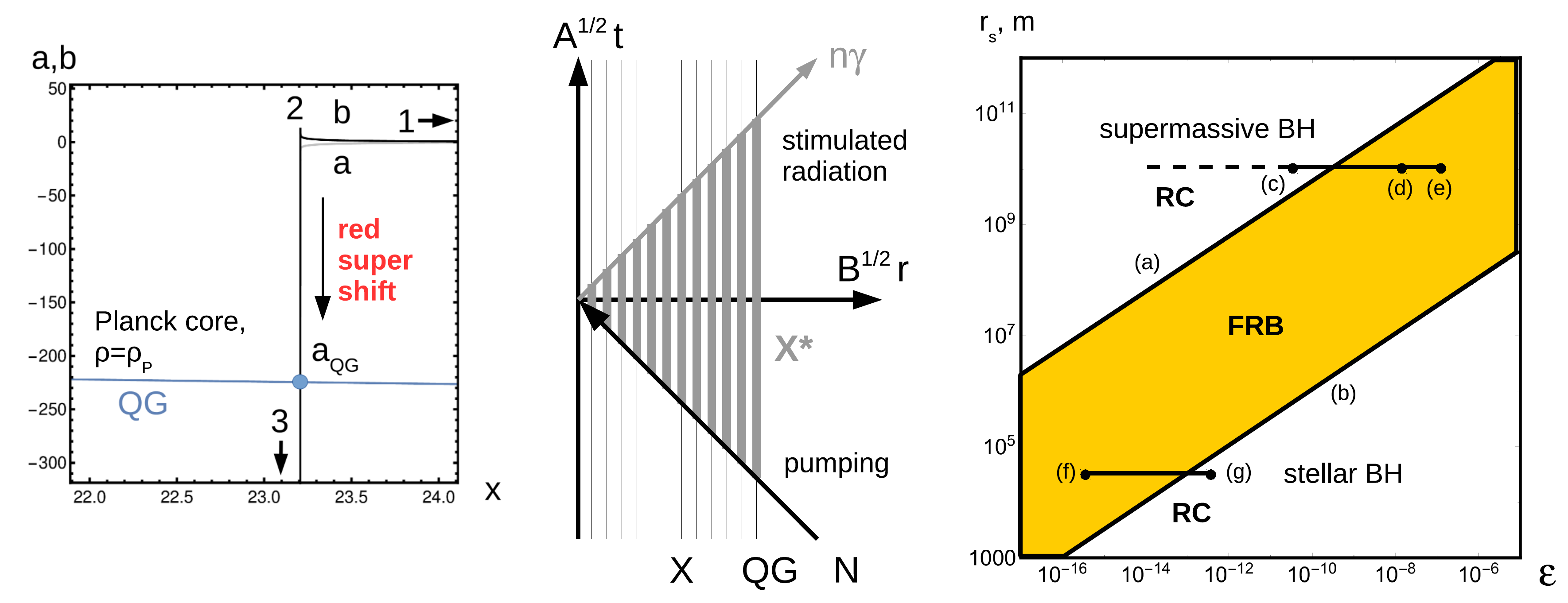}
\end{center}
\caption{On the left: quantum gravity cutoff in RDM-model for Milky Way scenario. In the center: a mechanism for generating FRB in RDM-model. On the right: simultaneous analysis of rotation curves and FRBs in RDM-model. Images from \cite{dm_stars}.}\label{p1f5}
\end{figure}

\paragraph*{Quantum gravity cutoff.} Further, we omit index {\it eff} in the formulae, assuming that the effective density and pressure are always considered. Also, for definiteness, we fix the dark matter to null type (NRDM). The resulting model is equivalent to a perfect fluid with the following equation of state (EOS):
\begin{eqnarray}
&&\rho=p_r,\ p_t=0,\label{rhopeff}
\end{eqnarray}
there is a relativistic relation between mass density and radial pressure, while the transverse pressure is switched off. The formulae (\ref{rhopeff}) become
\begin{eqnarray}
&&\rho=p_r=\epsilon/(8\pi r^2A).
\end{eqnarray}
Further, for illustration, we consider the solution for the Milky Way galaxy with a concentrated RDM-star in the center. Fig.\ref{p1f5} left shows the corresponding metric profiles. The solution starts in point 1 far away from the center, then in point 2 attempts to go to the Schwarzschild regime. We remind that the scale is logarithmic and the metric profiles jump many orders of magnitude near the point 2. Then they fall into abyss due to the red supershift phenomenon. Much earlier than the $A$-profile reaches minimum in point 3, the Planck density is achieved $\rho\sim\rho_P$. At this point we stop the solution and place a Planck core below it. Since the $B$-profile is also very small at this point and according to the formula (\ref{transf}), the Planck core possesses negative total mass, whose repulsive force supports the whole system in equilibrium. In further computations, only the order of the magnitude is important, on necessity, corrections can be applied via phenomenological factors \cite{dm_stars}. Taking into account that $\rho_P=l_P^{-2}$ in the units used, where $l_P$ is Planck length, also that redshift factor falls rapidly at almost constant $r\sim r_s$, the value in the cutoff point becomes
\begin{eqnarray}
&&A_{QG}=\epsilon\,(l_P/r_s)^2/(8\pi).\label{Aqg}
\end{eqnarray}
For the Milky Way, substituting the estimation of the $\epsilon$-parameter above and the known gravitational radius $r_s$ of the central black hole from Ghez et al. \cite{0808.2870}, we obtain $ A_ {QG} ^ {1/2} = 1.7 \cdot10 ^ {- 49} $. This value will be important for our further calculations.

\paragraph*{RDM-stars as sources of Fast Radio Bursts.} The common property for all dark star solutions is the presence of high energetic phenomena and strong redshift in their depths. Therefore, high energy photons created in these phenomena on the way out can be shifted to a long wave diapason. This makes dark stars natural candidates for sources of FRBs, the powerful flashes of extragalactic origin, registered in radio band. The lowest FRB frequency of 111~MHz has been reported by Fedorova and Rodin \cite{1812.10716}, the highest of 8~GHz -- by Gajjar et al. \cite{1804.04101}. Detailed experimental characteristics of FRBs can be found in {\it frbcat} catalogue by Petroff et al. \cite{frbcat}, there is also a catalogue of existing FRB theories {\it frbtheorycat} by Platts et al. \cite{frbtheorycat}. At the time of this writing, 118 distinct FRB sources have been registered and 59 FRB theories have been created.

A particular scenario with an RDM-star generating an FRB has been considered in \cite{dm_stars}. An object of an asteroid mass falls onto the RDM-star. The gravitational field inside the star acts as an accelerator with super-strong ultrarelativistic factor $\gamma=A_ {QG} ^ {-1/2}\sim10 ^ {49}$. The nucleons $N$ composing the asteroid enter in the inelastic collisions with particles $X$ forming the Planck core, producing the excited states of a typical energy $ E (X^*) \sim \sqrt {2 m_X E_N} $. The high-energy photons formed by the decay of $X^*$ with energy $E(\gamma,in)\sim E(X^*)/2$ are subjected to super-strong redshift factor $\gamma^{-1}=A_ {QG} ^ {1/2}\sim10 ^ {-49}$. The $\gamma$-factors do not compensate each other due to the presence of the square root in the formula. Thus, the outgoing energy $E(\gamma,out)\sim\sqrt{m_Xm_N /(2\gamma)}$, the wavelength $\lambda_{out} = \sqrt {2\lambda_X \lambda_N \gamma}$. Taking $\lambda_X\sim l_P$, we obtain a formula for FRB wavelength
\begin{eqnarray}
&&\lambda_{out}=2 (2\pi)^{1/4} \sqrt{\lambda_N r_s}\,/\epsilon^{1/4},\label{lamout2}
\end{eqnarray}
containing only Compton wavelength of nucleon $\lambda_N $ and $(r_s,\epsilon)$-parameters of the RDM-star. Interestingly, Planck values are canceled out of the formula. Further, taking $\lambda_N=1.32\cdot10^{-15}$m, $r_s=1.2\cdot10^{10}$m, $\epsilon=4\cdot10^{-7}$, the wavelength and frequency of FRB are
\begin{eqnarray}
&&\lambda_{out}=0.5\textrm{m},\ \nu_{out}=0.6\textrm{GHz},\label{lamout3}
\end{eqnarray}
that falls in the observed range 0.111-8GHz of FRB frequencies.

Further evaluations can be found in \cite{dm_stars}. A snowball mechanism is introduced for generating a sequence of excited states, which produces the energy spectrum of photons cut from above by the computed $E_{out}$ value. The spectrum is open towards low energy values, however, the increasing scatter broadening dilutes the signal there. A common mechanism of stimulated emission (aka LASER) can generate a short pulse of coherent radiation, by the scheme displayed on Fig.\ref{p1f5} center. Other parameters, such as pulse width and pulse delay, spectral and beam efficiency, as well as polarization, repetition and periodicity, observed for some FRBs, have been also discussed in \cite{dm_stars}. Most of these parameters are insensitive to the nature of the FRB source, being imparted by local environment and/or interstellar/intergalactic medium on the way of signal propagation. These parameters can be described by the known source independent astrophysical mechanisms, such as scatter broadening and signal dispersion, as well as scenarios with an FRB source passing through a planetary system or an asteroid belt.

The estimation above has been made for a simplified scenario with a concentrated RDM-star in the center of the galaxy with Milky Way alike parameters. Fig.\ref{p1f5} right shows more possibilities. The coordinates $(r_s,\epsilon)$ are the gravitational radius and the parameter defining the contribution of a particular black hole (=RDM-star) to the galactic dark matter halo, $\epsilon=GM\lambda/(c^2L_{KT}N)$, where $M,\lambda,L_{KT}$ are parameters from the fit of the galactic structures explained at the beginning of this section, $N$ is the number of black holes in the structure. The band shows detected FRB frequencies between the lines (a) and (b), according to (\ref{lamout2}). Two horizontal lines show two classes of solutions, supermassive and stellar black holes, according to the scenarios considered above: (c) s2, (d) s3, (e) corresponds to the minimal velocity value $v\sim100$km/s on the plots Fig.\ref{p1f2}, (f) $\epsilon=4\cdot10^{-7}$ divided to $N=10^9$ stellar black holes, (g) the same with $N=10^6$, the estimations of the number of stellar black holes are from Wheeler and Johnson \cite{1107.3165} and references therein.

The main conclusion from the analysis of the plot on Fig.\ref{p1f5} right is that the band and the horizontal lines have an intersection, therefore the model of RDM-stars is able to describe simultaneously the rotation curves and FRBs. Moreover, two solution classes exist, stellar and supermassive black holes. The plot is constructed on the basis of Milky Way data, extracted from its highly detailed rotation curve, and is valid for galaxies of similar structure. It would be interesting to populate it with data from other galaxies, that depends on the availability of rotation curves with a comparable detalization.

\section{TOV-stars with Planck core}

\begin{figure}
\centering
\includegraphics[width=\textwidth]{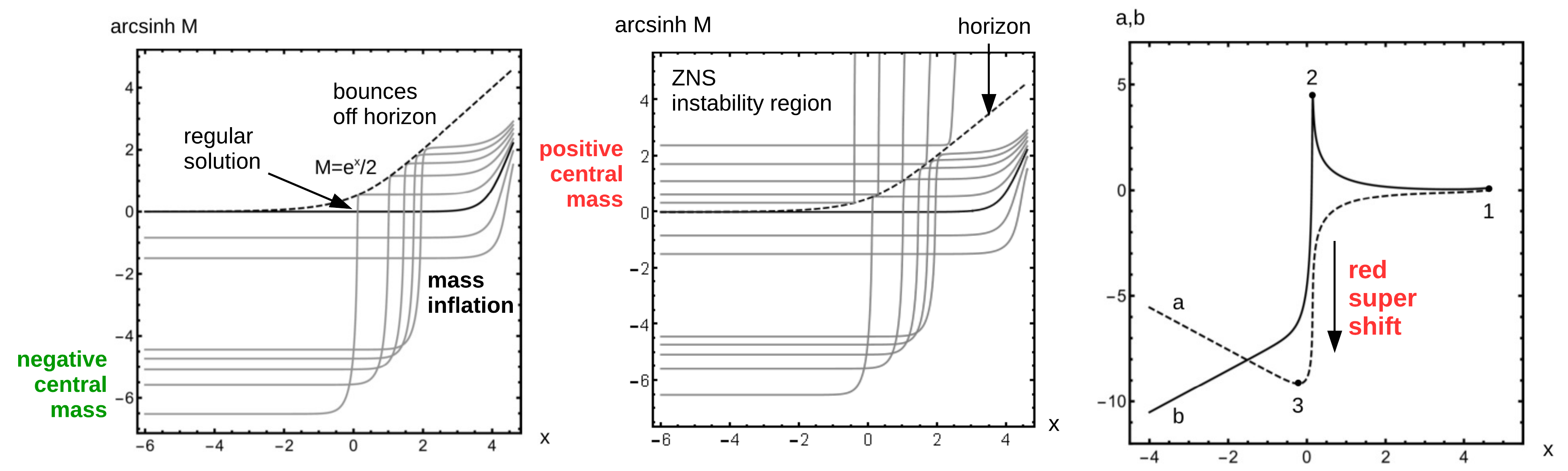}
\caption{TOV-star solutions. On the left: solutions with negative central mass, in the center: with positive central mass, with phenomenon of Zel'dovich-Novikov-Starobinskij explained, on the right: a typical behavior of metric coefficients.}
\label{p2f1}
\end{figure}

In this section we consider Tolman-Oppenheimer-Volkoff (TOV) stars. It is well known system, described by EOS
\begin{eqnarray}
&&w\rho=p_r=p_t,\label{toveos}
\end{eqnarray}
differing from RDM-stars EOS (\ref{rhopeff}) by the presence of two components of transverse pressure $T^\mu_\nu=\diag(-\rho,p_r,p_t,p_t)$, equally distributed with the radial one (isotropic pressure). Parameter $w$ regulates the composition and temperature of the star. Small values $w=kT/(mc^2)$ correspond to an ideal gas of massive particles of a given temperature. In this section we will mainly consider an ultrarelativistic plasma or photon gas, corresponding to the value $w=1/3$. The Einstein equations have a form (see, e.g., Blau \cite{Blau}):
\begin{eqnarray}
&&w \rho'_r=-(\rho M/r^2)(1+w)(1+4\pi r^3 w\rho/M)(1-2M/r)^{-1},\label{tov1}\\
&&M'_r=4\pi r^2\rho,\ h'_r=4\pi r(1-2M/r)^{-1}\rho(1+w),\label{tov2}
\end{eqnarray}
where the metric coefficients are chosen as
\begin{eqnarray}
&&A=e^{2h}f,\ B=f^{-1},\ f=1-2M/r.
\end{eqnarray}
A consequence of this system is so the called hydrostatic equation
\begin{eqnarray}
&&r (p+ \rho) A'_r + 2 A r p'_r =0,
\end{eqnarray}
possessing an analytical solution
\begin{eqnarray}
&&4\pi w\rho=k_1 A^{k_2}\label{hydrosol}
\end{eqnarray}
with constants
\begin{eqnarray}
&&k_1=4\pi\rho_1 w,\ k_2=-(1 + 1/w)/2,\ k_3=\log k_1.
\end{eqnarray}
The system can be rewritten in logarithmic variables (\ref{transf}), to the form convenient for a numerical solution:
\begin{eqnarray}
&&a'_x=-1 + e^b + 2 e^{2 x + k_2 a + b + k_3},\label{tovax}\\ 
&&b'_x= 1 - e^b + (2/w)e^{2 x + k_2 a + b + k_3}.\label{tovbx}
\end{eqnarray}
They are complemented by initial data $a_1=0$, $b_1=-\log(1-2M_1/r_1)$, where $w=1/3$, $k_2=-2$, $k_3=\log(4\pi\rho_1/3)$, $\rho_1$ and $M_1$ at a large $r_1$ are given. The typical solution is shown on Fig.\ref{p2f1} left. The coordinates are $x=\log r$ and $\arcsinh M$, the last one possesses asymptotics $\pm\log|2M|$ at large $|M|$, convenient to display all features of a solution in a single plot. 

Usually a regular solution is investigated, satisfying a condition $M=0$ in the center. This solution is shown by a thick line on the figure. We investigate what happens if this condition is relaxed. If a solution with the same $\rho_1$ is started with smaller $M_1$, below the regular line, it simply ends below this line in a negative central value. More interesting, if the solution is started above the regular line, it will not end in a positive central value and will not cross a horizon. Instead, it bounces off the horizon, goes through the mass inflation and ends in an even more negative central value. These solutions are clearly singular in the center, however, they are of interest to us, since the quantum gravity cutoff considered below can remove these singularities, replacing them with a regular Planck core.

For completeness we also consider a case of Fig.\ref{p2f1} center, when the solution is started above the horizon line, physically under the horizon. It similarly bounces off the horizon from inside and goes to the positive central value. If one reverts the integration, the solution started under the horizon from a positive mass Schwarzschild singularity will stay inside the horizon. This phenomenon was discovered by Zel'dovich, Novikov, Starobinskij \cite{ZNS} investigating the formation of white holes under the influence of matter ejected from the central singularity. The system is described by similar equations and the result is that the ejected matter never leaves the horizon and the white hole under the described circumstances cannot explode. This effect (internal ZNS instability) is one of instability types inherent to white holes, the other one (external Eardley instability) will be considered below in the Appendix. Mainly, in this paper, we consider a dual solution, possessing negative mass Schwarzschild singularity and evolving outside of the instability region.

The plot on Fig.\ref{p2f1} right shows the typical evolution of metric coefficients, appearing to be very similar to such plots for RDM-stars. An important difference is that the redshift fall and the mass inflation for a TOV-star appear to be much more moderate in comparison with an RDM-star of similar parameters. Table~\ref{p2tab1} shows the scenario with a stellar mass compact object in a cosmic microwave background, described by TOV equations. Although the variation of metric coefficients and enclosed mass in physical units is very large, it is still much smaller than the analogous variations for RDM-stars.

\begin{table}
\begin{center}{\footnotesize
\caption{TOV-star, scenario with a stellar mass compact object in cosmic microwave background}\label{p2tab1}

~

\def\arraystretch{1.1}
\begin{tabular}{|c|c|}
\hline
model parameters&$M_1=10 M_\odot$, $w=1/3$, $\rho_1=\rho_{cmb}=4\cdot10^{-14}$~J/m$^3$\\ \hline
starting point of&$r_1=10^6$m, $a_1=0$, $b_1=0.0299773$,\\ 
the integration& $M_1/M_\odot=10$ \\ \hline
&$r_{2}=29532.4$m, $a_{2}=-54.2719$, $b_{2}=53.7265$, \\ 
supershift begins&$M_2/M_1-1=-3.64729\cdot10^{-23}$,\\
& $r_2-2GM_2/c^2=1.37139\cdot10^{-19}$m\\ \hline
&$r_{3}=20638.1$m, \\
supershift ends&$a_{3}=-107.522$, $b_{3}=-104.685$,\\ 
& $\log_{10}(-M_3/M_\odot)=46.3087$\\ \hline
minimal radius& $r_4=1.62\cdot10^{-35}$m, \\ 
(Planck length),&$a_4=-17.6594$, $b_4=-195.278$,\\
end of the integration& $M_4=1.728\ M_3$\\ \hline
\end{tabular}

}\end{center}
\end{table}

\begin{table}
\begin{center}{\footnotesize
\caption{micro TOV-star, the critical case}\label{p2tab2}

~

\def\arraystretch{1.1}
\begin{tabular}{|c|c|}
\hline
model parameters&$w=1/3$, $\rho_1=\rho_{cmb}=4\cdot10^{-14}$~J/m$^3$, $a_{QG}=-146.264$\\ \hline
starting point of&$r_1=1.13042\cdot10^{-2}$m, $a_1=0$, $b_1=0.0100503$,\\ 
the integration& $M_1=7.61132\cdot10^{22}$kg \\ \hline
&$r_{2}=1.13042\cdot10^{-4}$m, \\ 
supershift begins&$a_{2}=-73.6529$, $b_{2}=73.0876$,\\
& $r_2-2GM_2/c^2=2.04973\cdot10^{-36}$m\\ \hline
&$r_{3}=7.89967\cdot10^{-5}$m, \\
supershift ends&$a_{3}=-146.264$, $b_{3}=-143.408$,\\ 
& $\log_{10}(-M_3/M_\odot)=54.7085$\\ \hline
minimal radius& $r_4=1.62\cdot10^{-35}$m, \\ 
(Planck length),&$a_4=-75.7824$, $b_4=-214.619$,\\
end of the integration& $M_4=1.72898\ M_3$\\ \hline
\end{tabular}

}\end{center}
\end{table}

Considering this scenario in more details, we see that $r_2-r_s\sim10^{-19}$m, the object comes very close to the gravitational collapse. This is a distance where the matter terms, initially weak, representing cosmic microwave background, are amplified and start to dominate in the equation. Although this is the result of a purely classical model, quantum considerations can change this number.

Further, $\log_{10}(-M_3/M_\odot)\sim46$, in comparison with the mass of the observable universe: $\log_{10}(M_{uni}/M_\odot)\sim23$. Thus, the considered compact object contains a core of negative mass, by absolute value much greater than the mass of the universe, compensated by the coat of TOV matter with almost the same positive mass. A similar computation for RDM-model gives an even larger number: $\log_{10}(-M_3/M_\odot)\sim10^5$. 

These enormous numbers could be the result of model idealization. Their origin is the unrestrained phenomenon of mass inflation. It can be changed if a (non-gravitational) interaction between the counterstreaming flows and corresponding corrections to EOS will be taken into account. Also, the considered solutions are stationary and can take enormous amount of time to form. A qualitative interpretation of the obtained solutions is that a permission of negative mass (Planck core) leads to a polarization of the solution to the parts with highly positive and highly negative masses, almost compensating each other in the result.

The other origin of large numbers is Planck density: $\rho_P=5\cdot10^{96}$kg/m$^3$. Straightforward estimation for the Planck density core of only $R=1$mm radius gives the mass $M=(4/3)\pi R^3\rho_P=2\cdot10^{88}$kg, gravitational radius $R_s = 2GM/c^2= 3\cdot10^{61}$m, much larger than the mass and the radius of the observable universe $M_{uni} = 10^{53}$kg, $R_{uni} = 4\cdot10^{26}$m. Such a core will immediately cover the universe by its gravitational radius, with a large margin. To place such objects in our universe, a mechanism for mass compensation is necessary. For instance, the one of this paper, effectively negative masses created by quantum gravity and coated by positive mass shells until the equilibrium with a moderate mass value is reached.

Enormous reserve of energy hiding inside TOV-stars can fuel extremely high-energetic phenomena. Figuratively speaking, if such a bubble bursts somewhere, the consequences can be felt throughout the universe. Thus, it is natural to consider these objects as potential sources of FRBs and we will do this, at first considering the quantum gravity cutoff and formation of Planck core in the center of TOV-star. 

\paragraph*{Quantum gravity cutoff.} Setting $w=1/3$ in a solution of TOV hydrostatic equation (\ref{hydrosol}), obtain $\rho\sim A^{-2}$. Therefore for the considered scenario with cosmic microwave background: $\rho_P/\rho_{cmb}=A_{QG}^{-2}$. Taking $\rho_P=4.633\cdot10^{113}$J/m$^3$ and $\rho_{cmb}=4.19\cdot10^{-14}$J/m$^3$ from Longair \cite{Longair}, in energetic units, have $A_{QG}=(\rho_{cmb}/\rho_P)^{1/2}=3\cdot10^{-64}$, $a_{QG}=-146$. The question now is whether such value can be reached. For RDM-stars with physically interesting parameters the redshift fall is enormous and the Planck density can be always reached before achieving the minimum in $a$-dependence. For a TOV-star, the redshift fall and associated density increase in solutions are moderate. The solutions shown in Fig.\ref{p2f1} right and Table~\ref{p2tab1} pass the minimum $a_3$ before reaching $a_{QG}$ and the Planck density for these solutions is not reached. The necessary condition for formation of the Planck core is $a_3<a_{QG}$. To investigate a satisfaction of this condition, we have performed the following numerical experiment. Keeping the outer density fixed to $\rho_{cmb}$, we changed the solution mass, or associated parameter $x_{10s}=\log_{10}r_s$, where $r_s$ is Schwarzschild radius in meters, in the range $x_{10s}\in[-10,10]$. After the integration of TOV-equations, we detected the minimum $a_3$ and found that it is well approximated by linear dependence $a_3=-128.089 + 4.60519\cdot x_{10s}$. As a result, $a_3<a_{QG}$ condition is satisfied at $r_s<r_{s,crit}=0.11$mm (micro TOV-stars), $M<7.6\cdot10^{22}$kg, approximately Moon's mass. The critical case is shown in Table~\ref{p2tab2}. The equality $a_3=a_{QG}$ and the resulting $r_2\sim r_s$ confirms this computation.

\paragraph*{TOV-stars as sources of Fast Radio Bursts.} Let us consider a photon of initially Planck energy, $E_{in}\sim E_P$, $\lambda_{in}\sim l_P$, on the surface of the Planck core. After applying the redshift, the outgoing wavelength $\lambda_{out}=l_P A_{QG}^{-1/2}=0.9$mm. Experimentally it is $\lambda_{exp}=37.5$mm, for the highest 8~GHz FRB detection of FRB121102 source \cite{1804.04101}. The deviation factor $\lambda_{exp}/\lambda_{out}\sim40$ can still be considered as a good hit, taking into account 127 orders of difference in the input density parameters $\rho_{cmb}/\rho_P$. Technically, it can be compensated by an attenuation factor $E_{in}= E_P/N$, the initial photon is $N\sim40$ times weaker than Planck energy. A part of this factor can be related with $(1+z)$ cosmological redshift of the source, $z\sim0.2-0.3$, the remaining factor to explain is $N\sim30$.

The analytical formula for the wavelength is also interesting: $\lambda_{out}=l_P(\rho_{cmb}/\rho_P)^{-1/4}$, or, in Planck units, simply $\lambda_{out}=\rho_{cmb}^{-1/4}$, depending only on the cosmic microwave background density.

Consideration of other FRB parameters proceeds similar to \cite{dm_stars}. Most of the parameters depend not on the source, but on its environment and propagation medium of the signal. Here we consider one question: can the bursts repeat? For the critical case $r_s=r_{s,crit}$ and isotropic estimation of the total burst energy from Cao et al. \cite{0193}, there is an inner reserve of energy for $7.6\cdot10^{22}\textrm{kg}\cdot c^2/(10^{32-34}\textrm{J})\sim10^{6-8}$ bursts. The energy can be also refilled from the environment, e.g., a companion, an asteroid belt, etc. In this refilling, when the threshold $r_s>r_{s,crit}$ is passed, the conditions for Planck core existence disappear. This can trigger the FRB, that will return the system to $r_s<r_{s,crit}$ state.

In summary, TOV-stars can also be the sources of FRB, or may represent a species of these signals. Differently to FRB from RDM-star, triggered by the fall of an asteroid, TOV-star signals can be autogenerated, possessing also a mechanism for autonomous oscillations around the critical state.

\paragraph*{Comparison of different models.} While both RDM- and TOV-stars can generate FRBs, the asymptotically flat rotation curves are generated only by RDM-stars. Only they possess the necessary $\rho\sim r^{-2}$ dependence, while the considered TOV solutions have $\rho\to Const>0$ asymptotics. 

On the other hand, in Barranco et al. \cite{1301.6785} a different solution of the TOV system has been investigated, possessing $\rho\sim r^{-2}$ dependence. It is a well known analytical self-similar solution, whose existence follows from scale-invariance of the system, see, e.g., the work by Visser and Yunes \cite{0211001}. Due to the appropriate density profile, this solution can be used to describe the rotation curves, a configuration known as isothermal dark matter halo. In addition, the asymptotic velocity on this solution appears to be $(v/c)^2=2 w/(1 + w)$. The experimentally observed velocities are non-relativistic, achievable only for small $w$. From here a conclusion is drawn, that the dark matter composing the galaxies ``must be cold''.

If one uses RDM model instead of TOV, a physically different system with the absence of transverse pressure is formed. Here all types of dark matter produce the same density profile $\rho\sim r^{-2}$ and the same asymptotically flat rotation curves. The value of the orbital velocity is defined by the parameter $(v/c)^2=\epsilon$, while the type of the matter by the other parameter $c_5$. As a result, the consideration of rotation curves in the RDM model does not impose a restriction on the type of dark matter in the galaxies.

Self-similar solutions of the TOV system form a very special class, different from the ones considered in this section. The regular type solutions we consider look like a ball of almost constant density, with a little bump of density in the center, due to self-gravitation. The singular solutions we consider have the same outer asymptotics, just possess a concentrated negative mass or a regular Planck core in the center. Self-similar solutions possess such a strong self-gravitation, that the whole solution shape is changed, also at large distances. This is possible only at a very large mass of solution. Especially, for the photon gas we consider, the mass should be enormous to make the light condense under its own gravitation. The computation shows $\rho_1^*=\epsilon^*/(4\pi r_1^2)$, $M_1^*=\epsilon^* r_1$, $\epsilon^*=2w/(1 + 6 w + w^2)$, in geometrical units, for self-similar solution. With $w=1/3$, at $r_1=3.1\cdot10^{21}$~m, the outer range of the Milky Way galaxy, it is $\rho_1^*=0.21$~J/m$^3$, $M_1^*=4.5\cdot10^{17}M_\odot$, being compared with $\rho_{cmb}=4\cdot10^{-14}$~J/m$^3$, $M_{cmb}=(4\pi/3)\rho_{cmb}r_1^3=2.8\cdot10^4 M_\odot$. Thus, the mass characteristics of the system we consider are 13 orders of magnitude below the formation of self-similar solutions.

The other question is an ability of Planck stars directly generate FRBs, investigated by Barrau, Rovelli, Vidotto in \cite{1409.4031}. The BRV model considers a collapse of primordial matter to a black hole going through the quantum bounce to the eruption of the white hole. The eruption appears at a delayed time due to strong gravitation. The time of recollapse depends on the mass of the star and is estimated to $t=0.2M^2$, in Planck units. Equating it with Hubble time, the mass and the size of Planck stars are estimated, created at the Big Bang and exploding ``today'': $M=(5t_H)^{1/2}=1.2\cdot10^{23}$kg, $r_s=2M=(20t_H)^{1/2}=0.2$mm. This estimation comes close to the critical size of TOV-stars $r_{s,crit}=0.11$mm obtained in our model.

The BRV model predicts an observable FRB signal at $\lambda\sim r_s\sim0.2$mm. The cosmological redshift correction can be also applied. The result is numerically similar to our model ($\lambda\sim0.9$mm), although obtained in a completely different setup: recollapse of Planck stars vs redshift of photons of initially Planck energy that arise in stationary TOV solutions with the Planck core in thermal equilibrium with CMB. Our prediction $\lambda\sim\rho_{cmb}^{-1/4}$ and the BRV formula $\lambda\sim(20t_H)^{1/2}$ coincide up to a numerical factor $\sim4.7$, if one takes into account cosmological constraints $\Omega_{cmb}=\rho_{cmb}/\rho_{crit}=4.2\cdot10^{-5}$, $\rho_{crit}=3H^2/(8\pi)$, $t_H=1/H$.

In the original BRV model of Planck stars only non-repeating FRBs are possible. The work by Barceló et al. \cite{1409.1501} proposes repeating recollapses and final stabilization of an object due to dissipative effects. Such a stationary object can be equivalent to the RDM- and TOV-stars discussed here. After its formation, it can produce both repeating and non-repeating FRBs depending on the environment.

The further paper by Barceló et al. \cite{1511.00633} seems to ``close'' the topic of Planck stars, referring to Eardley instability of the white hole part. Due to this instability, the white holes under the influence of external radiation would turn into black holes, not having time to emit the FRB. Below, in Appendix, we will bring a contra-argument, showing that Eardley instability can be eliminated if the core of the white hole possesses negative mass. Physically, it can be the Planck core, formed as the result of quantum gravity corrections when the Planck density is reached. Therefore the models from the Planck star family, as well as the FRB estimates based on them, avoid the white hole instabilities in a self-consistent way.

\section{Discussion: What is dark matter made of?}

In this section we consider three hypotheses on the composition of dark matter, based on the considered dark star models.

\paragraph*{Hypothesis 1:} the galactic dark matter can be cold, can be hot, producing the same rotation curves.

\vspace{3mm}
It follows from the solution properties of the RDM model, the orbital velocity depends only on intensity factor $\epsilon$, not on matter constitution (cold/hot, M/N/T cases, controlled by the other constant $c_5$). It can be a new type of particles, which can be sterile for interactions with the known matter sectors, i.e., enter only in gravitational interactions with them. It can be almost sterile, i.e., other interactions allowed at high energies in Planck cores, while extremely weak at low energies outside. 

One more fascinating possibility is that the dark matter is composed of known particles, placed in an unusual condition. Let consider a photon of Planck energy, emitted from the surface of the Planck core of an RDM-star: $E_{in}\sim E_P$, $\lambda_{in}\sim l_P$. Applying the redshift $ A_ {QG} ^ {1/2} \sim10 ^ {- 49} $, have $\lambda_{out}=l_P A_{QG}^{-1/2}\sim10^{14}$m. It is an extremely large wavelength, about 4 light days, 16 times the Sun-Pluto distance. Such longwave photons can not be registered by usual means, e.g., via radio telescopes. Although the energy of every such photon is extremely small, they come in numbers providing the necessary mass density to explain the rotation curves of the galaxies. The detailed consideration shows that at the Planck core one Planck energy particle per Planck area per Planck time is emitted, that corresponds to Planck density and pressure on its surface. After that the factor $ A_ {QG} ^ {1/2}$ is applied twice, for redshift and gravitational time dilation, then the geometrical $(r_s/r)^2$ factor corresponds to the measured halo density $\rho = p_r = 1/l_P^2\cdot A_{QG}\cdot (r_s/r)^2  = \epsilon/(8\pi r^2)$, where for the redshift factor the formula (\ref{Aqg}) is used.

In the considered scenario the particles should be massless. For massive particles the Compton length must be greater than $\lambda_{out}\sim10^{14}$m, obviously, excluding lightest neutrino species and other massive particles. Those particles do not overcome the gravitational barrier and remain bounded inside the RDM-star. From the Standard Model, the only appropriate particles for this scenario can be photons and gravitons. Scenarios with massive particles should have a larger starting energy to overcome the barrier.

\begin{figure}
\centering
\includegraphics[width=\textwidth]{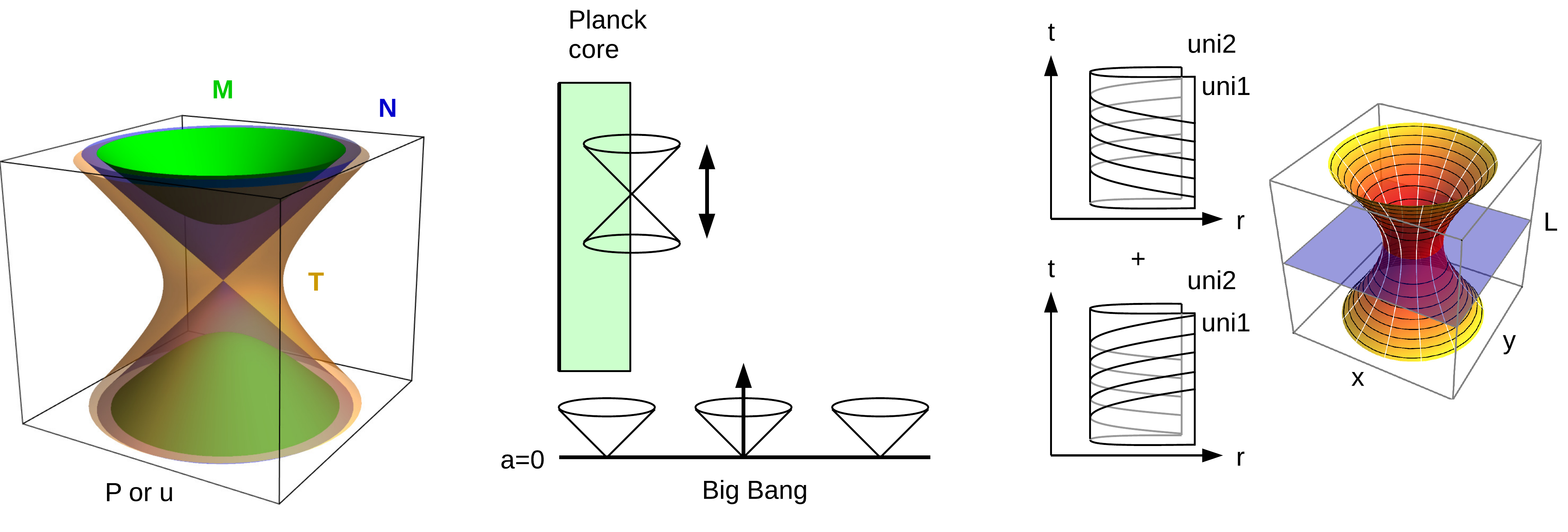}
\caption{Illustration to hypothesis~2. On the left: the mass shells for massive, null and tachyonic particles. In the center: a difference between Big Bang and Planck core light cones structure. On the right: the world lines of dark matter particles captured by a wormhole.}
\label{p3f1}
\end{figure}

\paragraph*{Hypothesis 2:} the emission of galactic dark matter from a Planck core is T-symmetric, in future and in past directions.

\vspace{3mm}
We remind that an RDM-star contains two T-symmetric flows, ingoing and outgoing ones. Fig.\ref{p3f1} left shows the mass shells for momentum $P$ or velocity $u$ vectors. There is a one-sheet tachyonic shell, containing both ingoing and outgoing directions, and two-sheet massive/null shells, where these directions are separated. In any case, we assume that all mass shells become completely occupied at the Planck core. The reason can be an extremely high temperature, in Planck range $T\sim T_P$, the one achievable at Big Bang. It is so hot there, that the vicinity of the Planck core becomes insensitive to the external thermodynamical time arrow and develops an own, T-symmetric thermodynamics. An important difference in this context is that RDM singularity and Planck core are timelike, while Big Bang singularity is spacelike. Different orientation of light cones can lead to the absent time arrow (recovered T-symmetry) near the Planck core and its presence near/after the Big Bang. This difference is shown in Fig.\ref{p3f1} center, the Big Bang light cones have only the upper part, while the Planck core light cones have both, T-symmetrically occupied parts.

One technical remark about $P_0<0$ parts of the mass shells. Although they formally correspond to negative energy and seem to be related with negative mass exotic matter, really they just correspond to T-conjugated flows of the same particles as $P_0>0$ counterparts. To verify this, consider T-reflection, that reverts $P^\mu$ and $u^\mu$ vectors, as well as orientation of the world lines, while preserves the action $A=m\int d\tau |x'_\mu x'^\mu|^{1/2}$ and the energy-momentum tensor $T^{\mu\nu}=\rho u^\mu u^\nu$, which are only physically important.

One more exotic possibility for T-symmetric emission is that the world lines of dark matter are captured by a geometry of wormhole, as shown on Fig.\ref{p3f1} right. The ingoing flows from one universe become outgoing in the other universe and vice versa. In a stationary scenario, their T-symmetric superposition can be chosen.

Independently on the detailed properties of the considered models, the emission of T-symmetric type is necessary due to simple physical reasons. If only outgoing flows would be present, the total mass of a dark matter halo could not be greater than the mass of (quasi) black holes, from where it originates. Experimentally, the halo mass is much greater than the mass of black holes. In the considered setup, ingoing and outgoing flows compensate each other and allow for arbitrary ratio between halo and black hole masses.

\begin{figure}
\centering
\includegraphics[width=\textwidth]{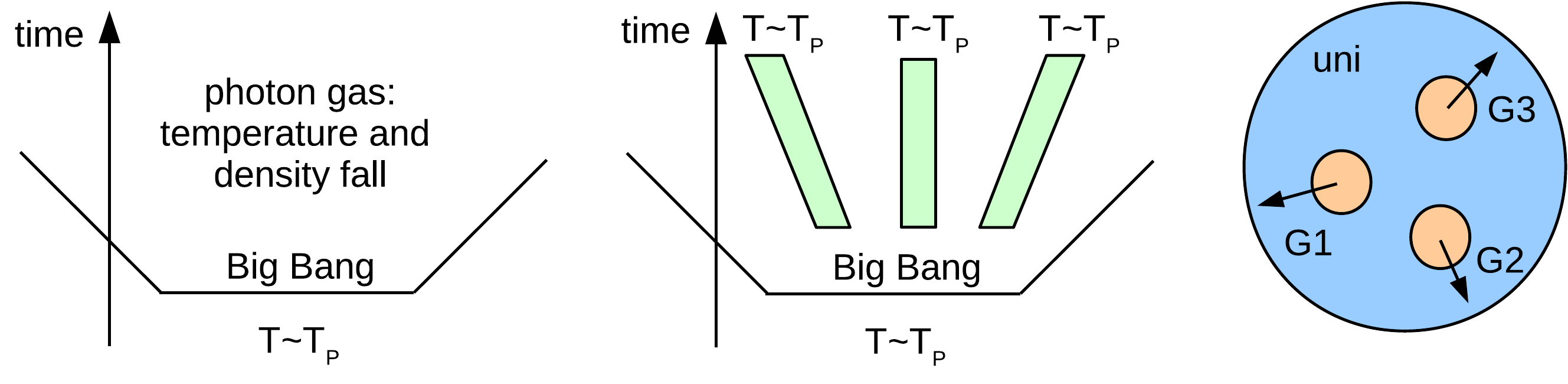}
\caption{Illustration to hypothesis~3. On the left: evolution of photon gas in standard cosmology. In the center: the same with RDM-stars. On the right: Swiss cheese model with galaxies filled by hot dark matter, surrounded by cold dark matter, in expanding universe.}
\label{p3f2}
\end{figure}

\paragraph*{Hypothesis 3:} the cosmological dark matter mimics cold type.

\vspace{3mm}
The common opinion is that dark matter both in the galaxies and in between them is cold, i.e., is composed of massive non-relativistic particles. The hot cosmological dark matter would lead to a different expansion rate of the universe. Let us consider an evolution of uniform photon gas in standard cosmology, as shown schematically on Fig.\ref{p3f2} left. There is an initial flash, then the energy and the number density of photons fall in the expanding universe. For cold dark matter only the density falls. This makes a difference to the evolution of the energy-momentum tensor. 

On the other hand, the distribution of dark matter photons in the RDM model is different, see Fig.\ref{p3f2} center. Their initial energy at Planck core is fixed: $E\sim E_P$, the exit energy is also fixed by the local $A_{QG}^{1/2}$ factor. If the resulting distribution will possess a constant temperature, then in the long-range evolution it will behave like cold dark matter.

The other possibility is that EOS of cosmological dark matter is not identical to the galactic one. There is a class of Swiss cheese models, where the galaxies and their halos do not change their size and structure under cosmological expansion and move as a whole. The cosmological expansion acts only on the level where the matter distribution can be considered as uniform. The galaxies coated in massive halos can behave like macro-particles of cold dark matter, as shown on Fig.\ref{p3f2} right. Internally they can be filled with hot radiation, externally produce the same gravitational fields as cold massive particles.

As we have mentioned earlier, while fitting the Milky Way rotation curve, the experimental data are compatible with the presence of a cut of dark matter halo at $R_{cut}\sim50$kpc. While this cut was taken in the model just phenomenologically, the physical mechanisms for it can be constructed. Two-phase distribution, with hot radial dark matter inside the galaxy joined to cold uniform dark matter outside, can be used. It resembles a known phenomenon of termination shock on the border of the Solar system, where the solar wind meets the interstellar medium. This can be modeled similarly in galactic scales. A suitable mechanism for the termination can be any kind of interaction of dark matter particles in the ingoing/outgoing flows and the outer medium. In particular, it can be an absorption or a scatter of longwave dark photons by the intergalactic medium. 

\section{Conclusion}

In this paper we have experimented with the insertion of Planck core in several earlier known astrophysical models. What becomes possible as a result of such modification:

(1)~RDM solutions can be properly continued to the strong field mode. These are stationary solutions describing black holes, coupled to the radial flows of dark matter. In weak fields, such configuration of dark matter can be used as a model of spiral galaxies possessing realistic rotation curves. In this model, the geometric dependence of the density on the distance $\rho\sim r^{-2}$, typical for the RDM configuration in a single center approximation, gives flat rotation curves, while assuming the coupling of all black holes in the galaxy to RDM, deviations of the rotation curves from the flat shape are also described. In strong fields, a peculiar phenomenon of erasing the event horizon occurs; instead, a spherical region of super-strong redshift is formed. This phenomenon is accompanied by the effect of mass inflation by Hamilton-Pollack, in a thin layer near the gravitational radius a very large positive mass is accumulated, approximately compensated by the negative mass of the Planck core. Outside, such an object, an RDM-star, is perceived as a Schwarzschild black hole of limited mass. 

(2)~when an external body, for example, an asteroid, falls on an RDM-star, a flash of high-energy photons occurs, then the super-strong redshift of the RDM-star moves the flash frequency to the radio band. This process can be considered as a mechanism for generating fast radio bursts. The calculations lead to the formula for the wavelength $\lambda_{out}=2(2\pi)^{1/4}(\lambda_N r_s)^{1/2}/\epsilon^{1/4}$, where $\lambda_N$ is the Compton wavelength of the nucleons that make up the asteroid, $r_s$ is the gravitational radius of the RDM-star, $\epsilon=(v/c)^2$ is the parameter determining the orbital velocity of stars $v$ in the galaxy. Evaluation with the parameters of the Milky Way galaxy gives the wavelength $\lambda_{out}=0.5$~m and the frequency $\nu_{out}=0.6$~GHz, in the range 0.111 ... 8~GHz for the observed FRB frequencies. 

(3)~the Tolman-Oppenheimer-Volkoff system of equations describing the equilibrium of isotropic matter, for the Planck core located in the center, also has an interesting structure of solutions. For the matter, we consider photon gas stitched with cosmic microwave background at infinity. As a result of the calculation, a stationary solution with the parameters of a micro black hole $r_s<0.11$~mm is obtained. With a smooth change in the external boundary condition (slow accretion of external matter), autonomous oscillations arise in the system, accompanied by self-generation of photonic flashes. In this model, a formula for the outgoing wavelength is $\lambda_{out}=l_P(\rho_{cmb}/\rho_P)^{-1/4}$, numerically $\lambda_{out}=0.9$~mm, $\nu_{out}=333$~GHz, that fall close to the observed FRB range. 

(4)~white holes become stable. We have examined the dynamics of white holes in the Ori-Poisson model and showed that the insertion of negative mass core eliminates both Eardley and Zel'dovich-Novikov-Starobinskij types of instability. 

Based on the considered models, in frames of this work, we have proposed three hypotheses about the composition of astrophysical dark matter. 

(1)~In galaxies, the dark matter can be cold or hot, massive, null or even tachyonic, producing the same rotation curves. Particularly, it can be composed of massless particles with initially Planck energy, finally redshifted to the extremely large wavelength $\lambda_{out}\sim10^{14}$m. More particularly, it can be composed of low energy photons with such wavelength. 

(2)~The emission of dark matter particles happens in a T-symmetric way, in future and in past directions. This allows to explain the abundance of dark matter in comparison with the masses of its sources. 

(3)~At cosmological distances, the dark matter behaves like it is cold. Several mechanisms for this behavior have been proposed.

\section*{Acknowledgements}
The author thanks the organizers and participants of the XXIII Bled Workshop ``What comes beyond the Standard models?'' for fruitful discussions. The author also thanks Kira Konich for proofreading the paper.

\footnotesize

\section*{Appendix: Stability of white holes}

The model by Ori-Poisson \cite{OriPoisson}, describing a white hole or, more precisely, its juncture with a black hole, can be considered as a simplification of the stationary models presented in this paper. Instead of continuous superposition of ingoing and outgoing shells of matter, there are precisely two shells, one ingoing, one outgoing. The model is not stationary, it has a white hole as the initial state and a black hole as the final state. The advantage of the model is the existence of an analytical solution. The system is usually considered unstable, but we will now show that it is not.

\begin{figure}
\centering
\includegraphics[width=\textwidth]{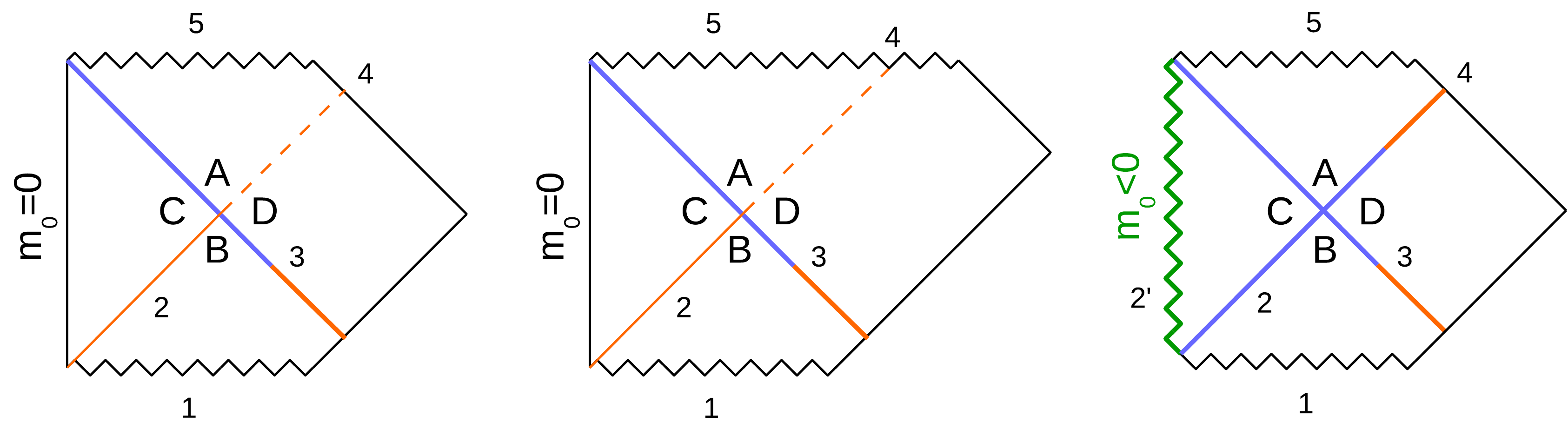}
\caption{Penrose diagrams for white holes in Ori-Poisson model. On the left: low efficient white hole eruption, in the center: no signal case, on the right: T-symmetric case.}
\label{p4f1}
\end{figure}

In addition to the previously described internal ZNS instability, there is external instability found in the work by Eardley \cite{Eardley}. In Ori-Poisson formulation, the instability can be illustrated by the Penrose diagram on Fig.\ref{p4f1} left. The original white hole (1) explodes, completely releasing its mass into an outgoing null shell (2). There is an ingoing null shell (3) of originally small energy. This shell cannot enter the white hole, in principle (like a shell that cannot exit from the interior of a black hole). Instead, it slows down at the horizon and after a long wait (one can substitute here Hubble time, for instance) receives a super-strong blue shift. It forms a thin super-energetic ``blue sheet''. Upon its collision with the outgoing shell, an analytically computable rearrangement of energy occurs. As a result, a negligible part of the initial energy comes out (4). The ingoing blue sheet disappears under the horizon of a newly formed black hole (5).

In Barceló et al. \cite{1511.00633} an even more asymmetric scenario is considered, depicted on Fig.\ref{p4f1} center. Here not only the ingoing, but also the outgoing shell disappears in the black hole. Absolutely nothing comes out of this system towards an external observer. 

One can immediately ask, how can it be that the black holes are stable, but the white holes are not? What about T-symmetry? As a resolution of this paradox, a T-symmetric scenario can be constructed, according to the principle ``for each incoming blue sheet, there is the same outgoing''. It is displayed on Fig.\ref{p4f1} right. Warning: the negative mass is required in the scenario. The original white hole (1) explodes, emitting more energy than its own mass in the form of an outgoing blue sheet (2). It leaves behind the core of negative mass (2'). After the collision between equal blue sheets (2) and (3), a shell of the same energy as the ingoing one comes out (4). The negative mass of the core is compensated by the incoming blue sheet, a black hole of the same mass as the original white hole is formed (5).

The details of the computation are the following. The solution consists of 4 Schwarz\-sch\-ild's patches, marked ABCD on the figures. Their masses are
\begin{equation}
m_A=M-E,\  m_B=M-dm,\  m_C=m_0,\  m_D=M,
\end{equation}
where $M$ is the mass of the white hole with the ingoing shell, $dm$ is the mass of the ingoing shell, $E$ is the mass/energy of the outgoing shell ($G = c = 1$), $m_0$ is a remainder. 

The computation is based on Dray-'t Hooft-Redmount (DTR) relation \cite{Dray-tHooft}:
\begin{equation}
f_Af_B=f_Cf_D,\ f_i=1-2m_i/R,\ i=A,B,C,D,
\end{equation}
where $R$ is the radius at shells intersection. Relative parameters are introduced:
\begin{equation}
\xi=R/(2M)-1,\ \alpha=dm/M,\ \beta=m_0/M,\ \eta=E/M,
\end{equation}
where $\xi$ measures the relative distance to the horizon, $\eta$ is the relative efficiency of white hole eruption. DTR relation results in the expression for the efficiency:
\begin{equation}
\eta=(1 - \alpha - \beta)\,\xi/(\alpha + \xi).\label{etares}
\end{equation}
As a side remark, in \cite{OriPoisson} a different global structure linking Schwarzschild's patches with a cosmological model was used, however, this does not influence the obtained efficiency formula. Now, let us consider the exponential evolution of the shell
\begin{equation}
\xi=\xi_0\exp(-\tau/(4M)).\label{xitau}
\end{equation}
Here $\tau=2t$ is the total time for the ingoing shell to reach the point of collision and for the outgoing shell to reach the distant observer, therefore a double factor in the formula. To bring in some values, for $r_s=1.2\cdot10^{10}$m it is an exponential process where the distance is halved every minute and for $\tau=13.8\cdot10^9$years the distance factor takes an enormously small value $\xi\sim\exp(-10^{16})$, practically insensitive to the starting $\xi_0$. 

Considering (\ref{etares}) for $\beta=0$, as in original Ori-Poisson paper, in the limit $0<\xi\ll\alpha\ll1$, obtain $\eta\sim\xi/\alpha\sim\exp(-10^{16})$, a vanishingly small efficiency of eruption.

Calculation with $\beta<0$ reveals a different class of solutions: $\beta\sim-\alpha/\xi$ results in $\eta\sim1$, 100\% efficiency, while for $\eta=\alpha$, $E=dm$, the T-symmetric case is reached selecting
\begin{equation}
\beta = -(\alpha^2 - \xi + 2 \alpha \xi)/\xi\sim-\alpha^2/\xi.
\end{equation}

The result of this computation demonstrates that in Ori-Poisson model Eardley instability can be eliminated if the system has a core of negative mass. Being combined with the earlier obtained result, both types of instability, Eardley and ZNS, can be removed from the white hole models by the introduction of a Planck core. A noticeable numerical difference between Ori-Poisson energetic parameters in comparison with those in the other considered models appears mainly due to the under-exponent Hubble time delays for the processes, that run continuously in the RDM and TOV models.

\end{document}